\documentclass[twocolumn,times]{aastex631}
\usepackage{showyourwork}
\usepackage{xspace}
\usepackage{amsmath}
\usepackage{amssymb}
\usepackage{bm}
\usepackage{physics}
\usepackage[super]{nth}
\usepackage{fontawesome5}
\usepackage{hyperref}

\newcommand{\edits}[1]{\textcolor{black}{#1}}

\begin{document}

\title{A Hierarchical Bayesian Framework for Inferring the Stellar Obliquity Distribution}

\correspondingauthor{Jiayin Dong}
\email{jdong@flatironinstitute.org}

\newcommand{\FlatironCCA}{Center for Computational Astrophysics, Flatiron Institute, 162 Fifth Avenue, New York, NY 10010, USA}

\author[0000-0002-3610-6953]{Jiayin Dong}
\altaffiliation{Flatiron Research Fellow}
\affiliation{\FlatironCCA}

\author[0000-0002-9328-5652]{Daniel Foreman-Mackey}
\affiliation{\FlatironCCA}

\received{May 1, 2023}
\revised{\today}
\accepted{}

\submitjournal{the AAS Journals}

\begin{abstract}

Stellar obliquity, the angle between a planet's orbital axis and its host star's spin axis, traces the formation and evolution of a planetary system. In transiting exoplanet observations, only the sky-projected stellar obliquity can be measured, but this can be de-projected using an estimate of the stellar obliquity.
In this paper, we introduce a flexible, hierarchical Bayesian framework that can be used to infer the stellar obliquity distribution solely from sky-projected stellar obliquities, including stellar inclination measurements when available. 
We demonstrate that while a constraint on the stellar inclination is crucial for measuring the obliquity of an individual system, it is not required for robust determination of the population-level stellar obliquity distribution.
In practice, the constraints on the stellar obliquity distribution are mainly driven by the sky-projected stellar obliquities.

When applying the framework to all systems with measured sky-projected stellar obliquity, which are mostly Hot Jupiter systems, we find that the inferred population-level obliquity distribution is unimodal and peaked at zero degrees. The misaligned systems have nearly isotropic stellar obliquities with no strong clustering near 90$\degr$.
The diverse range of stellar obliquities prefers dynamic mechanisms, such as planet-planet scattering after a convergent disk migration, which could produce both prograde and retrograde orbits of close-in planets with no strong inclination concentrations other than 0$\degr$.

\end{abstract}


\section{Introduction}
\label{sec:intro}

The stellar obliquity $\psi$ is the angle between a planet's orbital axis $\vb{\hat{n}_{\rm orb}}$ and the host star's spin axis $\vb{\hat{n}_\star}$. 
This angle is an important tracer of a planetary system's formation environment and dynamical evolution. The evolution of stellar obliquity can be roughly broken down into three stages. First, the formation and evolution of a protoplanetary disk determine the primordial stellar obliquity \citep[e.g.,][]{Bate10, Lai11, Batygin12}. Second, post-formation dynamical evolution in the planetary system, such as planet-planet scattering \citep[e.g.,][]{Rasio96, Chatterjee08, Nagasawa08, Beague12}, von Zeipel-Kozai-Lidov mechanisms \citep[e.g.,][]{Wu03, Naoz16}, and secular chaos \citep{Wu11}, can excite the mutual inclinations between planetary or stellar companions and alter the stellar obliquity. Lastly, the tidal force can reduce the stellar obliquity by realigning the host star's spin axis with the planet's orbital axis, if the tidal dissipation in the star is efficient \citep[e.g.,][]{Winn10, Albrecht12}. Additionally, massive stars with convective cores could generate internal gravity waves and dissipate angular momentum to their radiative zones, potentially affecting the stellar obliquity \citep{Rogers12, Rogers13}.

It is as yet unclear if, and to what extent, all of these physical and dynamic processes apply to exoplanetary systems. These proposed mechanisms all make different predictions on stellar obliquity distributions with a focus on Hot Jupiter systems \citep[see][and references therein]{Albrecht22, Dawson18}. For example, the secular chaos mechanism tends to produce a stellar obliquity distribution with $\psi < 90\degr$ \citep[e.g.,][]{Teyssandier19}. The stellar von Zeipel-Lidov-Kozai (ZLK) mechanism predicts a bimodal stellar obliquity distribution, concentrated at $40\degr$ and $140\degr$ \citep[e.g.,][]{Fabrycky07, Anderson16, Vick19}, assuming a zero stellar obliquity when the ZLK oscillation begins, or a broad stellar obliquity peaked near $90\degr$ if we drop the assumption \citep{Vick23}. The multiple-planet scattering mechanism results in a majority of aligned systems, with a small fraction of systems at a diverse range of stellar obliquities \citep[e.g.,][]{Beague12}. 
With these predictions in mind, we aim to determine the dominant mechanisms responsible for shaping close-in planetary systems by characterizing the stellar obliquity distribution of exoplanetary systems through a Bayesian approach.

When observing an exoplanet, typically only the sky-projected stellar obliquity $\lambda$, the angle between the projections of $\vb{\hat{n}_{\rm orb}}$ and $\vb{\hat{n}_\star}$ onto the plane of the sky, can be measured. This measurement is primarily obtained via the Rossiter-McLaughlin effect \citep{Rossiter24, McLaughlin24}. The stellar obliquity $\psi$ for an individual exoplanet system can be inferred, if both the sky-projected stellar obliquity $\lambda$ and the stellar inclination $i_\star$ are measured precisely. The relationship between $\psi$ and $\{\lambda, i_\star\}$ is given by \citep[e.g.,][]{Fabrycky09}:
\begin{equation}\label{eqn:psi}
    \cos{\psi} = \sin{i_\star}\sin{i_{\rm orb}}\cos{\lambda} + \cos{i_\star}\cos{i_{\rm orb}},
\end{equation}
where $i_{\rm orb}$ is the inclination angle between the vector $\bf{n}_{\rm orb}$ and the observer's line of sight, and $i_\star$ is the inclination angle between $\bf{n}_{\star}$ and the observer's line of sight.
If an exoplanet system transits, the orbit is nearly edge-on ($i_{\rm orb} \approx 90\degr$), so in those cases, this relationship becomes approximately
\begin{equation}
\cos{\psi} \approx \sin{i_\star}\cos{\lambda}\,,
\end{equation}
although this is not a simplification that we are required to make in this paper.

In some cases, stellar inclinations can be constrained via, for example, photometric and spectroscopic rotational modulation introduced by starspots for cool stars \citep[e.g.,][]{Masuda20, Albrecht21}, gravity darkening for fast-rotating stars \citep[e.g.,][]{Barnes09, Barnes11}, and asteroseismology for bright stars \citep[e.g.][]{Chaplin13}.
However, for the vast majority of exoplanet systems, $i_\star$ measurements are not feasible. In these cases, it is still possible to infer their stellar obliquities from the sky-projected obliquities, assuming isotropic stellar inclinations; however, the inferred $\psi$ will have greater uncertainty than the one inferred with $i_\star$ measurement \citep{Fabrycky09}.

The relationship between the distributions of stellar obliquity, sky-projected stellar obliquity, and stellar inclination is still not fully understood. In this study, we aim to gain a deeper understanding of this relationship and develop a statistical approach to infer the stellar obliquity distribution.
In Section~\ref{sec:coords}, we find the expression of sky-projected stellar obliquity $\lambda$ and stellar inclination $i_\star$ in terms of the orbital inclination $i_{\rm orb}$, stellar obliquity $\psi$, and the azimuthal angle of the stellar spin axis $\theta$.
In Section~\ref{sec:hbm}, we introduce a flexible, hierarchical Bayesian framework that allows us to infer the stellar obliquity distribution of a sample.
In Section~\ref{sec:simulation}, we examine the framework with simulated data and show that the inferred stellar obliquity distribution from sky-projected stellar obliquities is robust even if the $i_\star$ information is not provided.
Lastly, in Section~\ref{sec:observation}, we apply the framework to real observations and derive the stellar obliquity distribution for exoplanets. We discuss the implication of the stellar obliquity distribution for Hot Jupiter origins.

\section{Coordinate Setup and Transformation}\label{sec:coords}

\begin{figure*}[ht!]
    \script{coordinate.py}
    \gridline{
        \fig{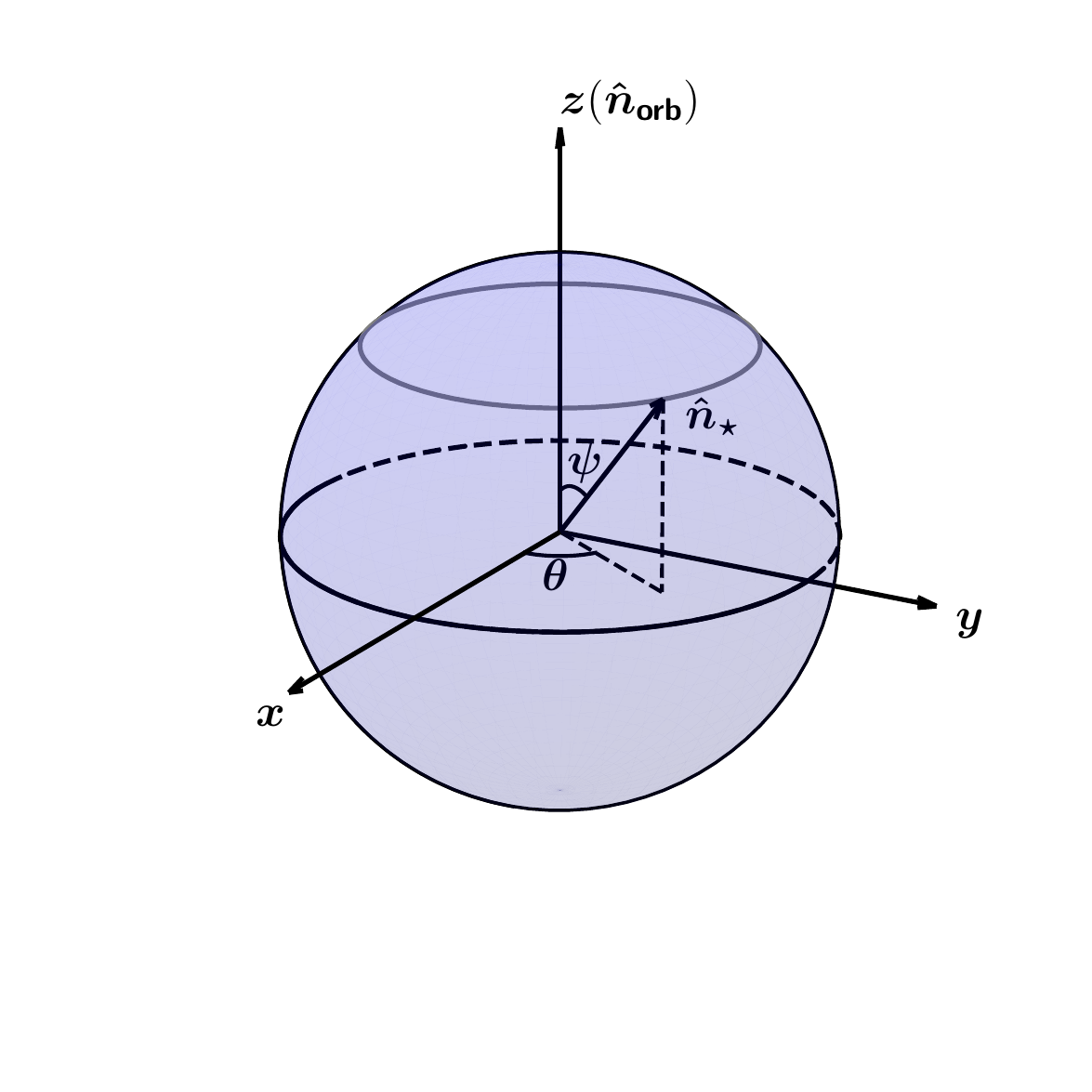}{0.45\textwidth}{\vspace*{-1.8cm}(a) The $\{\psi, \theta\}$ coordinate system. The planetary orbital axis is set to be aligned with $\vb*{z}$-axis. The grey circle corresponds to a constant $\psi$ value and its circumference is proportional to $\sin{\psi}$.}
        \fig{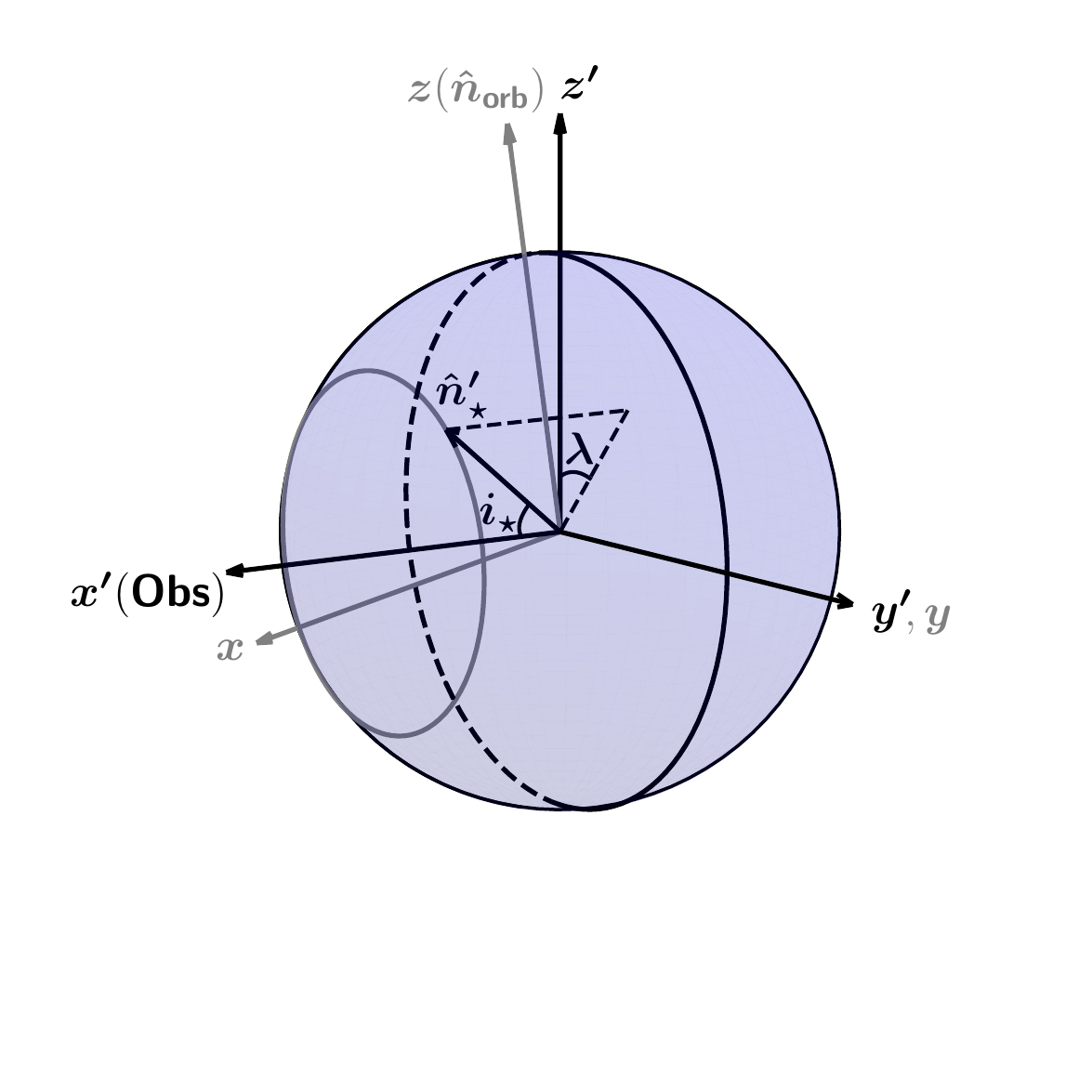}{0.45\textwidth}{\vspace*{-1.8cm}(b) The $\{\lambda, i_\star\}$ coordinate system. The observer's light of sight is set to be aligned with $\vb*{x'}$-axis. The grey circle corresponds to a constant $i_\star$ value and its circumference is proportional to $\sin{i_\star}$.}
    }
    \vspace*{-1cm}
    \caption{Two coordinate systems that describe the stellar spin axis $\vb{\hat{n}_\star}$ and the planet's orbital axis $\vb{\hat{n}_{\rm orb}}$. The $\{\psi, \theta\}$ coordinate system setup is motivated by the physical properties of a planetary system. The $\{\lambda, i_\star\}$ coordinate system setup is motivated by observables.}
    \label{fig:coord}
\end{figure*}

In this section, we find the expression of $\lambda$ or $i_\star$ in terms of stellar obliquity $\psi$, the azimuthal angle of the stellar spin axis relative to the orbital axis $\theta$, and the orbital inclination $i_{\rm orb}$. In Figure~\ref{fig:coord}, we introduce two coordinate systems that describe the stellar spin axis $\vb{\hat{n}_\star}$ and its planet's orbital axis $\vb{\hat{n}_{\rm orb}}$. The setup is similar to the coordinate system setup in \cite{Fabrycky09} but has a different definition of the azimuthal angle of the stellar spin and coordinate orientation.

The $\{\psi, \theta\}$ coordinate system, shown in the left panel in Figure~\ref{fig:coord}, is designed to describe the physical properties of a system. We set the planetary orbital axis $\vb{\hat{n}_{\rm orb}}$ as the $\vb{z}$-axis.
To define the stellar spin vector, we introduce the azimuthal angle of the vector around the orbital axis, $\theta$. 
The stellar spin axis $\vb{\hat{n}_\star}$ can be written in terms of $\psi$ and $\theta$ as
\begin{equation}
    \vb{\hat{n}_\star} = \sin{\psi}\cos{\theta}\vu*{x} + \sin{\psi}\sin{\theta}\vu*{y} + \cos{\psi}\vu*{z}.
\end{equation}

The $\{\lambda, i_\star\}$ coordinate system, shown in the right panel in Figure~\ref{fig:coord}, is designed to describe observables of a system. We set the observer's light of sight as the $\vb*{x'}$-axis.
Since only the difference between the sky-projected orbital axis and the sky-projected stellar spin axis can be measured, we conveniently align the projected orbital axis with the $\vb*{z}$-axis.
The stellar spin axis could be written in terms of $\lambda$ and $i_\star$ as 
\begin{equation}\label{eqn:nstar1}
    \vb{\hat{n}_\star}' = \cos{i_\star}\vu*{x}' + \sin{i_\star}\sin{\lambda}\vu*{y}' + \sin{i_\star}\cos{\lambda}\vu*{z}'.
\end{equation}

To transform the stellar spin vector $\vb{\hat{n}_\star}$ from the $\vb*{xyz}$ coordinate to the $\vb*{x'y'z'}$ coordinate, we rotate the $\vb*{xyz}$ coordinate by an angle $\pi/2 - i_{\rm orb}$ about the $\vb{y'}$-axis (the same as $\vb{y}$-axis). Conventionally, the inclination angle of the orbital axis $\vb{\hat{n}_{\rm orb}}$ is assumed to be less than $90\degr$, as what we adopt here.
Applying the rotation matrix such that
\begin{equation}
      \vb{\hat{n}_\star}' = \mqty[\sin{i_{\rm orb}} & 0 & \cos{i_{\rm orb}} \\ 0 & 1 & 0 \\ -\cos{i_{\rm orb}} & 0 & \sin{i_{\rm orb}}] \vb{\hat{n}_\star},
\end{equation}
we find the expression of $\vb{\hat{n}_\star}'$ in the $\vb*{x'y'z'}$ coordinate in terms of $\psi$, $\theta$, and $i_{\rm orb}$:
\begin{equation}\label{eqn:nstar2}
    \begin{split}
    \vb{\hat{n}_\star}' = (\sin{\psi}\cos{\theta}\sin{i_{\rm orb}}+\cos{\psi}\cos{i_{\rm orb}})\vu*{x}' &\\
    + \sin{\psi}\sin{\theta}\vu*{y}'& \\
    + (-\sin{\psi}\cos{\theta}\cos{i_{\rm orb}}+\cos{\psi}\sin{i_{\rm orb}})\vu*{z}'&.
    \end{split}
\end{equation}

Equating Equation~(\ref{eqn:nstar2}) and Equation~(\ref{eqn:nstar1}), we find the expression of $\lambda$ or $i_\star$ in terms of $\psi$, $\theta$, and $i_{\rm orb}$.
First, from the $\vb{\hat{x}'}$ terms, we get
\begin{equation}
    i_\star = \cos[-1](\sin{\psi}\cos{\theta}\sin{i_{\rm orb}}+\cos{\psi}\cos{i_{\rm orb}}).
\end{equation}
Next, dividing the $\vb{\hat{y}'}$ terms by the $\vb{\hat{z}'}$ terms, we get
\begin{equation}
    \lambda = \tan[-1](\frac{\sin{\psi}\sin{\theta}}{-\sin{\psi}\cos{\theta}\cos{i_{\rm orb}}+\cos{\psi}\sin{i_{\rm orb}}}).
\end{equation}
These two relations will be used in the hierarchical Bayesian framework to infer the stellar obliquity distribution.

\section{Hierarchical Bayesian Framework}\label{sec:hbm}

To find the stellar obliquity distribution of exoplanetary systems, we develop a hierarchical Bayesian framework that takes measurements of the observed sky-projected stellar obliquity $\lambda$ and orbital inclination $i_{\rm orb}$ as input data.
If a measurement of the stellar inclination $i_\star$ is available, it can be provided or inferred from the stellar rotation period $P_{\rm rot}$, stellar radius $R_\star$, and sky-projected rotational broadening velocity $v\sin{i}_\star$.
In the absence of an $i_\star$, $P_{\rm rot}$, or $v\sin{i}_\star$ measurement, the stellar obliquity $\psi$ distribution is inferred without $i_\star$ likelihoods.

\begin{figure}[ht!]
    \script{graph.py}
    \centering
    \includegraphics[width=0.8\linewidth]{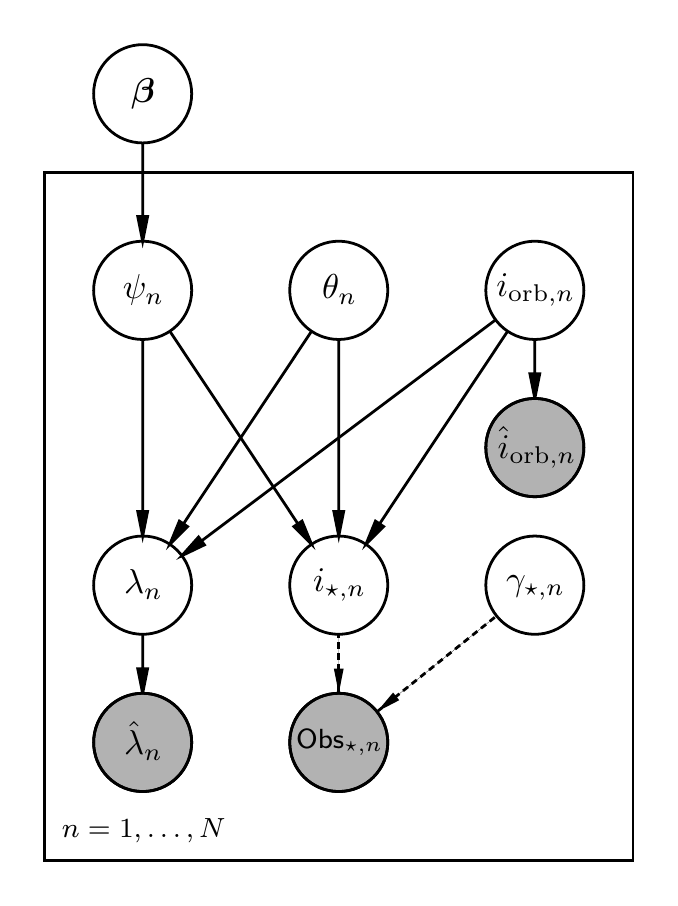}
    \caption{A probabilistic graphical model showing the conditional structure of our hierarchical Bayesian framework for inferring the stellar obliquity distribution of exoplanetary systems. The population model is parameterized by hyperparameters $\vb*{\beta}$, and it is constrained by the stellar obliquity $\psi_n$ of each $n_{\rm th}$ individual system. The stellar obliquity $\psi_n$, azimuthal angle $\theta_n$, and orbital inclination $i_{{\rm orb}, n}$ are constrained by the sky-projected stellar obliquity $\lambda_n$, and, if available, the stellar inclination $i_{\star, n}$.
    Properties of the star other than its inclination, such as its radius and rotation period, are represented by $\gamma_{\star, n}$, and these can inform the constraint on $i_{\star, n}$, if available. 
    $\hat{\lambda}_n$ has the measured sky-projected stellar obliquity and its uncertainty, Obs$_{\star,n}$ contains all observed properties of the star with uncertainties, and $\hat{i}_{{\rm orb}, n}$ has measured orbital inclination and its uncertainty.}
    \label{fig:graph}
\end{figure}

Figure~\ref{fig:graph} illustrates the probabilistic graphical model for our hierarchical Bayesian framework. We aim to constrain a set of hyperparameters $\bm{\beta}$ that describe the stellar obliquity distribution. The parameter set $\bm{\beta}$ is constrained by $N$ individual systems, where each $\psi_n$ is simultaneously fit based on the entire sample of sky-projected stellar obliquities $\lambda_n$, and if available, the stellar inclination $i_{\star, n}$. 
The parameter $\gamma_{\star, n}$ contains all stellar properties other than $i_{\star,n}$, such as the stellar rotation period $P_{{\rm rot},n}$, radius $R_{\star,n}$, and projected rotational velocity $v\sin{i_{\star,n}}$, if they are known.

In Figure~\ref{fig:graph}, the observed values of $\lambda_n$ and $i_{{\rm orb},n}$ are indicated by $\hat{\lambda}_n$ and $\hat{i}_{{\rm orb},n}$, respectively.
The $\hat{\lambda}_n$ measurements typically come from the Rossiter-McLaughlin effect or gravity darkening, and the $\hat{i}_{{\rm orb},n}$ is constrained by the transit light curve.

The constraints on stellar inclination are somewhat more complicated.
In Figure~\ref{fig:graph}, the Obs$_{\star,n}$ node includes any observed data that directly constrains $i_{\star,n}$.
This could include a direct measurement of $\hat{i}_{\star,n}$ \citep[via gravity-darkening or asteroseismology, for example][]{Barnes09, Chaplin13}, or a measurement of the sky-projected stellar rotational line broadening $\hat{v}\sin{i_{\star,n}}$, which is related to $i_\star$ as $v\sin{i_\star} = 2 \pi R_\star / P_{\rm rot}$ \citep{Masuda20}.

For the stellar obliquity distribution, we model the $\cos{\psi}$ distribution instead of $\psi$ distribution to understand if the stellar obliquity is isotropically distributed. If stellar obliquity distribution is isotropic, $\cos{\psi}$ is uniformly distributed between $-1$ and $1$.
To flexibly model the stellar obliquity distribution, we select a multi-component mixture of Beta distributions with hyperparameters $\bm{\beta} = \{\bm{w},\bm{\mu},\bm{\kappa}\}$, where each hyperparameter has a dimension of the number of components. For example, for a two-component mixture model, each hyperparameter has a dimension of 2.
We have $\bm{w} = \{w_0, w_1\}$, $\bm{\mu} = \{\mu_0, \mu_1\}$, and $\bm{\kappa} = \{\kappa_0, \kappa_1\}$.
The hyperparameter $\bm{w}$ describes the weight of each component. The $\bm{\mu}$ and 1/$\bm{\kappa}$ correspond to the mean and variance of each Beta distribution component, respectively, a reparametrization of parameters suggested by \cite{Gelman14}. The greater value of the $\kappa$, the smaller the variance (i.e., the distribution is more concentrated). The relations between the $\mu$ and $\kappa$ and the standard $\alpha$ and $\beta$ parameters in the Beta distribution are $\alpha = \mu \kappa$ and $\beta = (1-\mu) \kappa$.
This mixture distribution has the capacity to capture anything from an isotropic distribution to a strongly bimodal population.
Then, under this two-component model, the probability density function for $\cos{\psi}$ is
\begin{align}
    \edits{w_{0,1}} &\edits{\sim {\rm Bernoulli}(1/2)} \nonumber\\
    u_{0,1} &\sim {\rm Beta}(\mu_{0,1}\kappa_{0,1}, (1-\mu_{0,1})\kappa_{0,1}) \nonumber\\
    \cos{\psi} &= 2\,(w_0 u_0 + w_1 u_1) - 1.
\end{align}
Since the Beta distribution is defined on the interval $[0, 1]$ whereas the support of $\cos{\psi}$ is from $-1$ to $1$, we extend the mixture distribution's support $[0,1]$ to $[-1,1]$ using a linear transformation (i.e., $2\,( w_0 u_0 + w_1 u_1)-1$).
For hyperparmeters $\mu$ and $\kappa$, we adopt the following priors:
\begin{align}
    \mu_{0,1} &\sim \mathcal{U}(0, 1) \nonumber\\
    \log{\kappa}_{0,1} &\sim \mathcal{N}(0, 3),
\end{align}
where $\mu_{0,1}$ is uniformly distributed between $0$ and $1$ and $\log{\kappa}_{0,1}$ is normally distributed with a mean of $0$ and standard deviation of $3$. To deal with label switching in the mixture model, we remove the symmetry by forcing the vector $\vb*{\mu}$ to be ordered. 
Notably, when applying the framework to a small sample size with $N \lesssim 50$, the choice of hyperpriors for the Beta distribution could impact the inferred distribution \citep[e.g.,][]{Nagpal22, Gelman14}. To ensure the robustness of the inferred distributions in such cases, it is crucial to test their sensitivity to different hyperpriors.
\edits{We also note that our framework is flexible to customize to distributions other than the Beta distribution for the population-level stellar obliquity distribution inference. For example, if instead studying the stellar obliquity distribution in the $\psi$-angle space, a von Mises distribution could be used to evaluate the mean value and the dispersion of the angle.}

Next, the model parameter priors are the following:
\begin{align}
    \theta_n &\sim \mathcal{U}(0, \pi)  \nonumber\\
    \cos{i}_{{\rm orb},n} &\sim \mathcal{U}(0, 1).
\end{align}
If $\gamma_{\star, n}$ is available, we construct Normal distributions with means and standard deviations from measurements.
For the orbital inclination $i_{\rm orb}$, following the convention, we limit it to $\left[0, \pi/2\right]$, \edits{i.e., the orbital axis always pointing to us}. To not underestimate the $\psi$, we then set $i_\star$ to be varied from $0\degr$ to $180\degr$ \edits{such that the stellar spin axis could either point to us or point away from us}.
Besides, since $\lambda$ and $-\lambda$ correspond to the same $\psi$ solution, we limit $\lambda$ to $\left[0, \pi\right]$ and thus $\theta$ to $\left[0, \pi\right]$ \edits{to avoid bimodal distributions of $\lambda$ and $\theta$.}
We find doing so greatly improves the sampling performance while not compromising the inference of the $\psi$ distribution due to the symmetry. The $\left[0, \pi\right]$ support avoids the otherwise discontinuity at $\theta = 0$ and $\theta = \pi$.

The likelihood functions follow:
\begin{align}
    \mathcal{L}(\lambda) &\sim \prod_{n=1}^N\mathcal{N}(\hat{\lambda}_n, \sigma_{\hat{\lambda}_n}) \nonumber\\
    \mathcal{L}(i_\star) &\sim \prod_{n=1}^N\mathcal{N}({\rm Obs}_{\star,n}, \sigma_{{\rm Obs}_{\star,n}}) \,(\textit{optional})\nonumber\\
    \mathcal{L}(i_{\rm orb}) &\sim \prod_{n=1}^N\mathcal{N}(\hat{i}_{{\rm orb},n}, \sigma_{\hat{i}_{{\rm orb},n}}).
\end{align}

The probabilistic model is constructed using the $\mathtt{PyMC}$ package version $\mathtt{v5.1.2}$ \citep{pymc}, and the posteriors are sampled with the No-U-Turn Sampler \citep[NUTS;][]{Hoffman11}, which is a gradient-based Markov chain Monte Carlo (MCMC) sampling algorithm. This paper's figures and simulations are completely reproducible and were created using the $\mathtt{showyourwork}$ package. The open-source code is available on GitHub \href{https://github.com/jiayindong/obliquity}{(https://github.com/jiayindong/obliquity\,\faGithub)}.

\begin{figure}
    \centering
    \script{sim_plot.py}
    \includegraphics{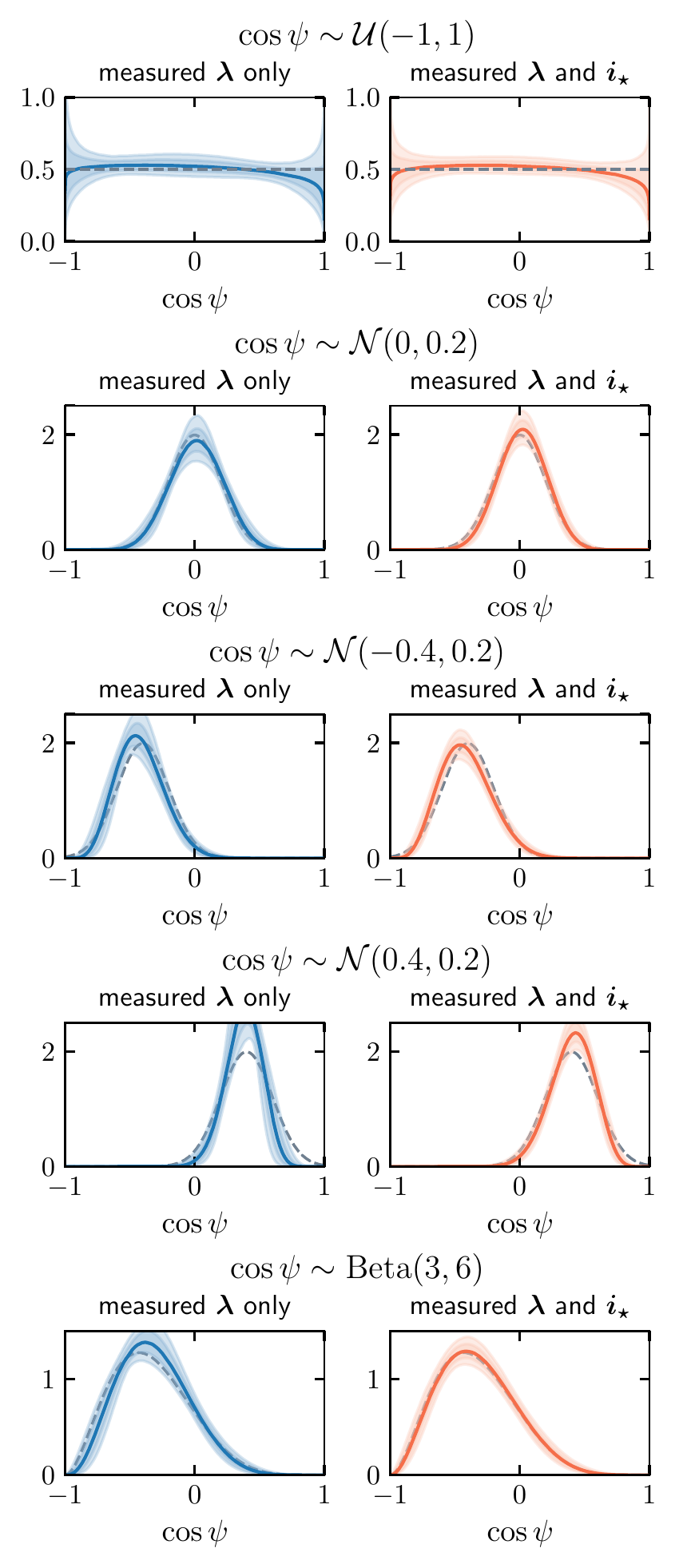}
    \caption{Inferred stellar obliquity distributions from sky-projected stellar obliquities \edits{without and with} information on the stellar inclination, depicted in blue and orange curves, respectively. Each row presents a set of simulated data with the true distribution of $\cos{\psi}$ indicated by grey dashed curves. The shallow contours represent the 1-$\sigma$ and 2-$\sigma$ uncertainties of the inferred distributions.}
    \label{fig:simulation}
\end{figure}

\begin{figure*}
    \centering
    \script{transform.py}
    \includegraphics[width=0.7\linewidth]{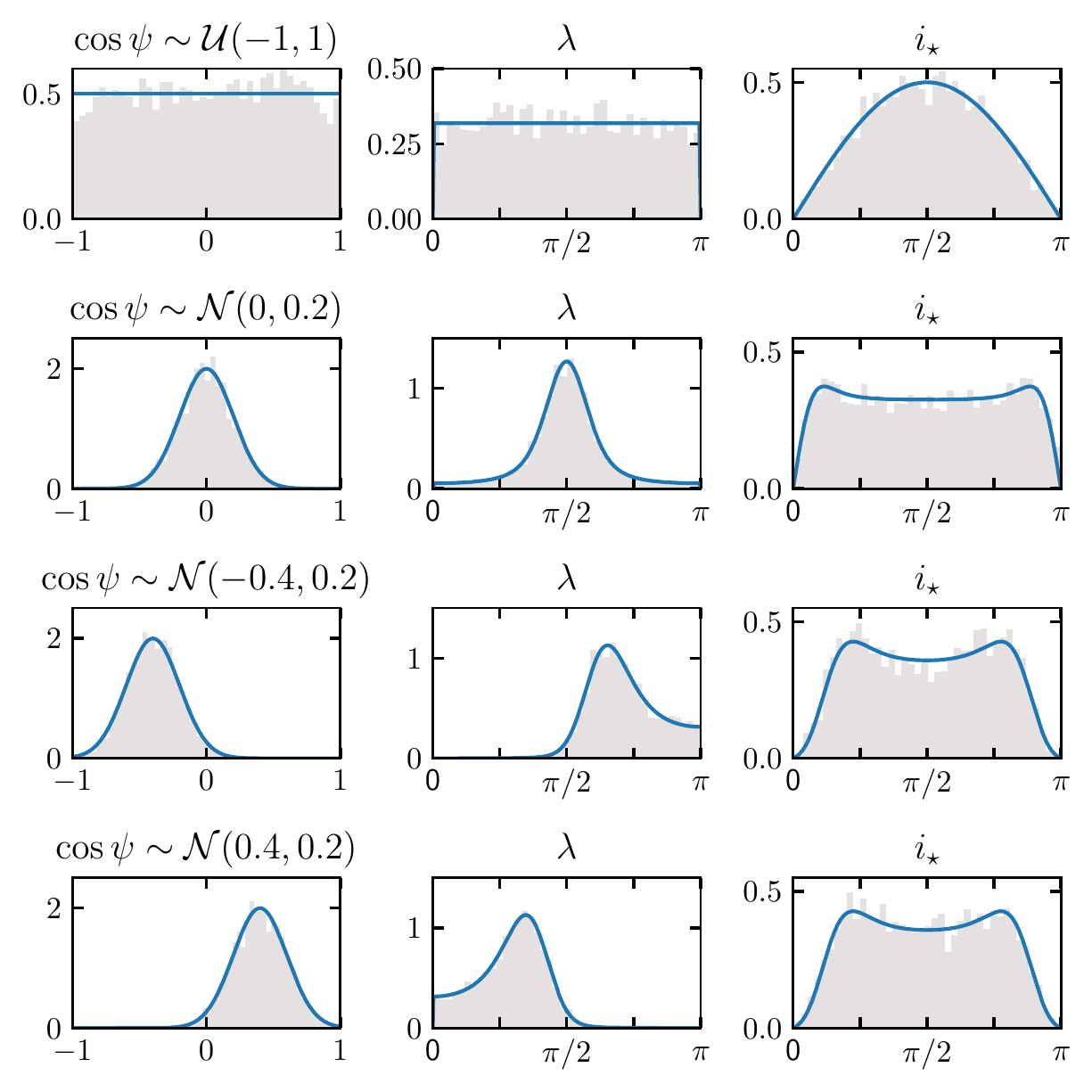}
    \caption{Simulated $\cos{\psi}$ distributions (\nth{1} column) and the corresponding distributions of sky-projected stellar obliquity $\lambda$ (\nth{2} column) and stellar inclination $i_\star$ (\nth{3} column). The grey histograms present the random samplings of $\lambda$ and $i_\star$ from the $\cos{\psi}$ distributions, and the blue curves present the numerical solutions.}
    \label{fig:transform}
\end{figure*}

\section{Model Performance}\label{sec:simulation}

\subsection{Simulated Data}\label{subsec:sim}

To investigate the performance of our hierarchical Bayesian framework, we apply it to simulated data, generated with a known ground-truth stellar obliquity distribution.
We test the following \edits{five} $\cos{\psi}$ distributions: a uniform distribution bounded between $-1$ and $1$ such that $\cos{\psi} \sim \mathcal{U}(-1,1)$, three truncated Normal distributions bounded between $-1$ and $1$, such that $\cos{\psi} \sim \mathcal{N}(0, 0.2)$, $\cos{\psi} \sim \mathcal{N}(-0.4, 0.2)$, and $\cos{\psi} \sim \mathcal{N}(0.4, 0.2)$, where the first number is the mean of the distribution and the second number is the standard deviation, \edits{and lastly, a Beta distribution such that $\cos{\psi} \sim \mathrm{Beta}(3, 6)$ with the support being extended linearly from $[0, 1]$ to $[-1, 1]$}.
For each $\cos{\psi}$ distribution, we randomly generate 200 samples of sky-projected stellar obliquity $\lambda$ and stellar inclination $i_\star$. We assume the stellar spin axis is uniformly distributed around the planetary orbital axis in the azimuthal direction and the orbital inclination is $90\degr$.
The sampled $i_\star$ and $\lambda$ here are \emph{true} values. 
To simulate the observation process, we add Gaussian noise to the \emph{true} $\lambda$ and $i_\star$, using uncertainties of $\sigma_{\lambda} = 8\degr$ and $\sigma_{i_\star} = 10\degr$, which are typical of the literature sample \cite{Albrecht22}.
Using these simulated $\lambda$ measurements and their uncertainties, we infer the $\cos{\psi}$ distribution of the sample with both $\lambda$ and $i_\star$ likelihoods or $\lambda$ likelihood only.

In Figure~\ref{fig:simulation}, we present the results of this experiment, plotting the inferred stellar obliquity distributions, compared to the ground-truth distributions.
Since the simulated stellar obliquity distributions only have a single component, we model the data with a single Beta distribution.
Each row of Figure~\ref{fig:simulation} corresponds to a different simulation distribution. The \edits{blue} curve and contours in the left column are the median, 1-$\sigma$ and 2-$\sigma$ uncertainties of the inferred $\cos{\psi}$ distribution \edits{when constraints on stellar inclination are not included}.
The right column shows the same inferences (in \edits{orange}) \edits{with stellar inclination information}.
Surprisingly, Figure~\ref{fig:simulation} demonstrates that our inference procedure recovers the true distribution for $\cos{\psi}$ equally well, regardless of the inclusion of $i_\star$ measurements.
Despite the fact that the inferred distributions without $i_\star$ measurements have marginally wider uncertainties, as indicated by shallow color contours, the modes and widths of the inferred stellar obliquity distributions are consistent with or without $i_\star$ likelihood.
Since the injected distributions in rows 2--4 are Normal distributions, it should not be surprising that the inferred distributions, which are Beta distributions, may not exactly match the injected distributions. Given that the true distributions are not included within the support of the distributions we are using to fit, it is impressive to see how well the underlying distributions could be recovered.

We also examine the role of orbital inclination $i_{\rm orb}$ in the stellar obliquity distribution inference. Since our study focuses on transiting-exoplanet systems, we consider an isotropic orbital inclination distribution between $80\degr$ and $90\degr$. This broad range of inclinations corresponds to an impact parameter range from 0 to 1 with a planet-star separation $a/R_\star$ of 6.
We compare the stellar obliquity distributions obtained by approximating $i_{\rm orb}$ to $90\degr$ with the distributions obtained using the actual $i_{\rm orb}$. We find the difference between the two distributions is negligible. This suggests that for transiting-exoplanet systems, approximating orbital inclinations as $90\degr$ will not compromise the stellar obliquity distribution inference.

We demonstrate through simulations that the inferred stellar obliquity distribution is robust even if the $i_\star$ information is not provided, and the $i_\star$ measurement only improves the constraint on the stellar obliquity distribution.

\subsection{Jacobian Transformations Between the $\psi$, $\lambda$, and \lowercase{$i_\star$} Distributions}\label{subsec:jacobian}

Next, we aim to gain insight into the reasons behind the predominant role of the sky-projected stellar obliquity distribution and the less significant impact of the stellar inclination distribution in the inference of the stellar obliquity distribution. To simplify the problem, the orbital inclination of the transiting planet is assumed to be $90\degr$ in this illustration. As discussed earlier, the assumption will not compromise the stellar obliquity distribution inference for transiting planets.
We pair the Cartesian components in Equation~(\ref{eqn:nstar1}) and Equation~(\ref{eqn:nstar2}), assuming $i_{\rm orb} = 90\degr$, and get:
\begin{align}
    \sin{\psi}\cos{\theta} = \cos{i_\star}& \label{eq:coord1}\\
    \sin{\psi}\sin{\theta} = \sin{\lambda}\sin{i_\star}& \label{eq:coord2}\\
    \cos{\psi} = \cos{\lambda}\sin{i_\star} \label{eq:coord3}&.
\end{align}

First, we derive the $\lambda$ distribution for a given $\cos{\psi}$ distribution. We could find the distribution of $\cos{\lambda}$ using the Jacobian transformation from $\cos{\psi}$ and $\cos{\theta}$. Since $\psi$ and $\theta$ are assumed to be independent variables, we could marginalize over $\theta$ to find the relation between the probability density functions between $\lambda$ and $\psi$.
The Jacobian transformation follows
\begin{equation}
    p(\cos{\lambda}) = \int \abs{\pdv{\cos{\psi}}{\cos{\lambda}}} p(\cos{\psi}) p(\cos{\theta}) d\cos{\theta}.
\end{equation}
Here $p(x)$ denotes the probability density distribution of $x$.
To find $\psi$ in terms of $\lambda$ and $\theta$, we replace $\sin{i_\star}$ in Equation~(\ref{eq:coord3}) using Equation~(\ref{eq:coord1}) and find
\begin{equation}
    \cos{\lambda} = \frac{\cos{\psi}}{\sqrt{1-(1-\cos^2{\psi})\cos^2{\theta}}}.
\end{equation} 
Reorganize the equation, we get 
\begin{equation}
    \cos^2{\psi} = \frac{\cos^2{\lambda}\cos^2{\theta}-\cos^2{\lambda}}{\cos^2{\lambda}\cos^2{\theta}-1}
\end{equation}
and the partial derivative 
\begin{equation}
    \abs{\pdv{\cos{\psi}}{\cos{\lambda}}} = \frac{(1-\cos^2{\theta})^{1/2}}{(1-\cos^2{\theta}\cos^2{\lambda})^{3/2}}.
\end{equation}
Since $\theta$ is uniformly distributed between $0$ and $\pi$, $p(\cos{\theta}) = p(\theta) \abs{\dv*{\theta}{\cos{\theta}}} = 1/\pi/(1-\cos^2{\theta)^{1/2}}$. Putting all the parts together, we get
\begin{equation}\label{eqn:jac_lam}
    p(\cos{\lambda}) = \frac{1}{\pi} \int_{-1}^{1} (1-\cos^2{\theta}\cos^2{\lambda})^{-3/2} p(\cos{\psi}) d\cos{\theta}.
\end{equation}
In the special case that $\cos{\psi}$ is uniformly distribution, i.e., $p(\cos{\psi}) = 1/2$, Equation~(\ref{eqn:jac_lam}) becomes $p(\cos{\lambda}) = 1/\pi/\sqrt{1-\cos^2{\lambda}}$, which is equivalent to $\lambda \sim \mathcal{U}(0, \pi)$. This suggests $\lambda$ is uniformly distributed for an isotropic $\psi$ distribution, as expected.

Next, we derive the $i_\star$ distribution for a given $\cos{\psi}$ distribution. Similarly, we first find the Jacobian transformation of $i_\star$ from $\psi$ and $\theta$ and then marginalize over $\theta$. It is easier to work on $\cos{i_\star}$ than $i_\star$:
\begin{equation}
    p(\cos{i_\star}) = \int \abs{\pdv{\sin{\psi}}{\cos{i_\star}}} p(\sin{\psi}) p(\cos{\theta}) d\cos{\theta}.
\end{equation}
From Equation~(\ref{eq:coord1}), we get 
\begin{equation}
    \sin{\psi} = \frac{\cos{i_\star}}{\cos{\theta}}
\end{equation} and the partial derivative
\begin{equation}
    \abs{\pdv{\sin{\psi}}{\cos{i_\star}}} = \frac{1}{\cos{\theta}}.
\end{equation} 
Again, we assume $\theta$ is uniformly distributed, and this gives $p(\cos{\theta}) = 1/\pi/(1-\cos^2{\theta)^{1/2}}$. Lastly, we transform the $p(\sin{\psi})$ to $p(\cos{\psi})$,
\begin{equation}
    p(\sin{\psi}) = \frac{2p(\cos{\psi})\sin{\psi}}{\sqrt{1-\sin^2{\psi}}}.
\end{equation}
The factor of 2 is from two solutions of $\cos{\psi}$ to the $\cos^2{\psi} = 1-\sin^2{\psi}$. Combining all the pieces together, we get
\begin{align}\label{eqn:jac_istar}
    p(\cos{i_\star}) = \nonumber\\ 
    \frac{2}{\pi} \int_{\cos{i_\star}}^{1,-1}& \frac{\cos{i_\star}/\cos{\theta}}{\sqrt{\cos^2{\theta}-\cos^2{i_\star}}}
    \frac{1}{\sqrt{1-\cos^2{\theta}}} p(\cos{\psi}) d\cos{\theta},
\end{align}
where the integral is from $\cos{i_\star}$ to $1$ for $\cos{\theta} > 0$, or from $\cos{i_\star}$ to $-1$ for $\cos{\theta} \leqslant 0$.
Note that the lower limit of the integral is $\cos{i_\star}$ instead of 0 since $\lvert \cos{\theta}/\cos{i_\star} \rvert \geqslant 1$.
If $\cos{\psi}$ is uniformly distribution, i.e., $p(\cos{\psi}) = 1/2$, the integral gives $1$, which suggests the $\cos{i_\star}$ is uniformly distributed, as expected.

Using Equation~(\ref{eqn:jac_lam}) and (\ref{eqn:jac_istar}), we can now derive the $\lambda$ and $i_\star$ distributions for any given $\psi$ distributions, assuming the azimuthal angle of the stellar spin axis $\theta$ is random. In Figure~\ref{fig:transform}, we present numerical solutions of the $\lambda$ and $i_\star$ distributions for four different $\cos{\psi}$ distributions used in Section~\ref{subsec:sim}. The top row of Figure~\ref{fig:transform} shows an isotropic $\psi$ distribution, where $\cos{\psi} \sim \mathcal{U}(-1,1)$. The second, third, and fourth rows of Figure~\ref{fig:transform} present truncated Normal distributions of $\cos{\psi}$ following $\mathcal{N}(0,0.2)$, $\mathcal{N}(-0.4,0.2)$, and $\mathcal{N}(0.4,0.2)$, respectively.
The blue curves in each row show the numerical solutions of the $\lambda$ and $i_\star$ distributions, while the grey histograms show the sampling of $\lambda$ and $i_\star$ from the $\cos{\psi}$ and $\theta$ distributions. 
For a uniform $\cos{\psi}$ distribution, the $\lambda$ distribution is uniform, and the $i_\star$ distribution is isotropic, proportional to $\sin{i_\star}$, as expected.

Interestingly, the $\lambda$ distribution is closely related to and sensitive to the underlying $\psi$ distribution, as demonstrated in the \nth{1} and \nth{2} columns in Figure~\ref{fig:transform}. For different stellar obliquities, the $\lambda$ distributions are distinguishable, making it possible to infer the $\psi$ distribution from the $\lambda$ distribution. On the other hand, the $i_\star$ distributions are less dependent on the underlying $\psi$ distribution. Compared to an isotropic $i_\star$ distribution, the curvature of the $i_\star$ distributions for different $\psi$ distributions differ the most at the low $i_\star$ values (i.e., $i_\star < \pi/4$), which places a challenge to observational detections. 
Additionally, the degeneracy of the solution could be a significant issue when attempting to infer the $\psi$ distribution from the $i_\star$ distribution. For example, when $\cos{\psi} \sim \mathcal{N}(-0.4,0.2)$ or $\cos{\psi} \sim \mathcal{N}(0.4,0.2)$, two $i_\star$ distributions are exactly the same, since the corresponding $i_\star$ distributions are identical for $\psi$ distributions symmetric around $\psi = \pi/2$.

The $\psi$ distribution can be inferred from the $\lambda$ distribution without loss of information due to the strong dependency of the $\lambda$ distribution on the $\psi$ distribution. It is also worthwhile to note that although we could find a mathematical expression of $\psi$ with $\lambda$ and $\theta$, the $\psi$ distribution cannot be inferred from the two variables since they are not independent variables.

\section{Application to Exoplanetary Systems}\label{sec:observation}

\begin{figure}
    \script{psi_plot.py}
    \begin{centering}
        \includegraphics{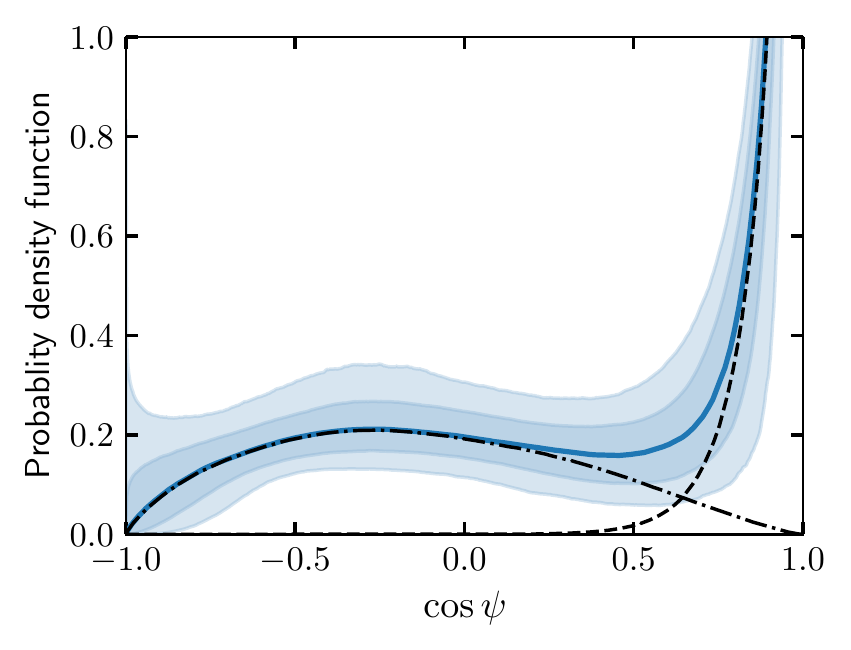}
        \caption{Inferred stellar obliquity distribution for all exoplanetary systems with sky-projected stellar obliquity measurements. This inference is based purely on the observed sky-projected obliquities using a two-component model. The blue line and shallow contours represent the median, 1-$\sigma$ and 2-$\sigma$ uncertainties of the inferred distributions. The medians of the two components are labeled in dashed and dotted-dashed lines, respectively.}
        \label{fig:psi_dist}
    \end{centering}
\end{figure}

\begin{figure*}
    \script{polar_plot.py}
    \begin{centering}
        \includegraphics{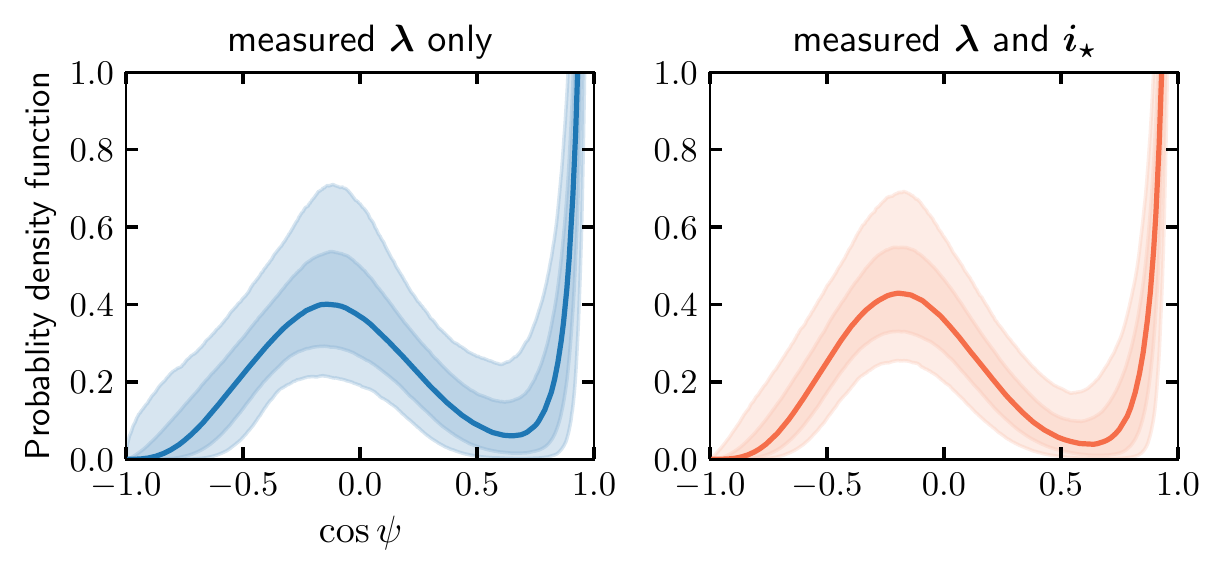}
        \caption{Inferred stellar obliquity distribution for the subsample of exoplanetary systems with both sky-projected stellar obliquity measurements and stellar inclination measurements. Inference made purely from measured projected stellar obliquity and inference made from both measured projected stellar obliquity and stellar inclination are shown in blue and orange curves, respectively. The line and shallow contours represent the median, 1-$\sigma$ and 2-$\sigma$ uncertainties of the inferred distributions.}
        \label{fig:polar}
    \end{centering}
\end{figure*}

We next apply our hierarchical Bayesian framework to a sample of 161 exoplanetary systems with sky-projected stellar obliquity measurements, primarily consisting of Hot Jupiter systems, as summarized in \cite{Albrecht22} Table A1.
\edits{We use a two-component model here for its flexibility in describing both an aligned-system population and a misaligned-system population that may or may not be concentrated at a certain angle.}
The inferred $\cos{\psi}$ distribution is shown in Figure~\ref{fig:psi_dist}. The $\cos{\psi}$ distribution is peaked at 1, with nearly flat behavior between $-0.75$ and $0.75$ and no significant clustering. The distribution suggests that there is a pile-up of planetary systems with stellar obliquities less than $40^\circ$ and an isotropic distribution for obliquities between $40^\circ$ and $140^\circ$.
The fraction of aligned systems $w_1$ dominates the distribution, with $w_1 = 0.719 \pm 0.085$. The corresponding posteriors for the parameters of the population Beta distribution are $\mu_1 = 0.976 \pm 0.022$ and $\kappa_1 = 14.1 \pm 26.6$. On the other hand, the fraction of misaligned systems is estimated to be $w_0 = 0.281 \pm 0.085$, with posteriors of $\mu_1 = 0.434 \pm 0.088$ and $\kappa_1 = 4.2 \pm 5.6$. Note that the $\mu_{0,1}$ needs to be transformed by $2\mu-1$ to represent to true means of the $\cos{\psi}$ distributions.

The discrepancy between this inference for the full sample of exoplanet systems and the previous analysis of the subsample with $i_\star$ measurements \citep{Albrecht21} warrants further investigation.
This earlier study identified a concentration of perpendicular planets and disfavored an isotropic stellar obliquity distribution. \edits{Indeed, when we applied our framework to this subsample, a concentration near $\cos{\psi} = -0.2$ was found in models including or without including the stellar inclination information, as shown in Figure~\ref{fig:polar}.}
There are at least two potential explanations for this difference: (1) the subsample with $i_\star$ measurements is small and only includes about 20 misaligned systems, and (2) the requirement for $i_\star$ measurements could introduce selection biases in the sample. 
First, since the sample size of the misaligned sample with $i_\star$ measurements is relatively small ($N < 30$), the observed sample, even if it is unbiased, may not be able to represent the underlying distribution. \edits{The small sample size leads to a large uncertainty on the inferred stellar obliquity distribution, shown by the 1-$\sigma$ and 2-$\sigma$ contours in Figure~\ref{fig:polar}.}
Second, the requirement of $i_\star$ measurements could introduce selection biases. The rotation modulation technique is most applicable to cool stars with spots, whereas misaligned Hot Jupiters are mostly found around hot stars. To measure the stellar inclinations of the hot-star hosts of misaligned Hot Jupiters, the gravity darkening technique is commonly used, but the technique is biased at detecting polar-orbit planets.

The inferred stellar obliquity distribution indicates that approximately $72\pm9$\% of the systems have a stellar obliquity of less than $40\degr$, and approximately $28\pm9$\% of the systems follow a nearly isotropic stellar obliquity distribution ranging from $\sim$$40\degr$ to $\sim$$140\degr$. These findings could have significant implications for the formation and evolution of close-in planetary systems, particularly on Hot Jupiters. The diverse distribution disfavors dynamical mechanisms, such as secular chaos which tends to produce stellar obliquities less than $90\degr$, or stellar Kozai which tends to produce stellar obliquities clustered at certain angles.
The broad distribution of misaligned systems is in good agreement with the predicted outcome of multiple giant planets scattering after a convergent disk migration, as proposed by various studies, such as \cite{Nagasawa11} and \cite{Beague12}. The intriguing result should be further examined with a more carefully selected sample of Hot Jupiters and provides opportunities to place constraints on their origin channels of Hot Jupiters.

\section{Summary \& Discussion}

In this work, we demonstrated that the stellar obliquity distribution could be robustly inferred from sky-projected stellar obliquities purely.
We introduced a flexible, hierarchical Bayesian framework for the stellar obliquity distribution inference. Stellar inclination measurements are optional input in the model, and if not available, they are assumed to be isotropically distributed.
Our open-source hierarchical Bayesian model, available on GitHub \href{https://github.com/jiayindong/obliquity}{(https://github.com/jiayindong/obliquity\,\faGithub)}, can be customized to different stellar obliquity distributions and priors for specific target samples.

It is crucial to consider the representativeness of the $i_\star$ sample when jointly modeling the stellar obliquity distribution from two data sets, one with and one without $i_\star$ measurements. An unrepresentative $i_\star$ sample could tighten the constraints on the stellar obliquities and bias the interpretation of the overall distribution.

Finally, we applied the framework to all exoplanetary systems to all exoplanetary systems with available sky-projected stellar obliquities and found that approximately $72\pm9$\% of the systems have a stellar obliquity less than $40\degr$, and approximately $28\pm9$\% of the systems follow a nearly isotropic stellar obliquity distribution between $\sim$$40\degr$ and $\sim$$140\degr$.
The distribution could have important implications for the formation and evolution of close-in planetary systems and is worth further investigation.

\section*{Acknowledgments}
We are grateful to Josh Winn and Jared Siegel for their valuable insights and feedback on this project.

This study was conducted using the \href{https://github.com/showyourwork/showyourwork}{$\mathtt{showyourwork}$} reproducible workflow \citep{Luger2021}, which leverages continuous integration to automate the data retrieval from \href{https://zenodo.org/}{zenodo.org}, figure generation, and manuscript compilation.
The script used to produce each figure can be accessed via a link in the corresponding figure caption, as it corresponds to the latest build of the manuscript.
The git repository for this study, which includes the Jupyter notebook demonstration and case studies of the stellar obliquity distribution inference, is publicly accessible at \url{https://github.com/jiayindong/obliquity}.

\vspace*{5mm}

\software{$\mathtt{ArviZ}$ \citep{arviz_2019}, $\mathtt{Jupyter}$ \citep{Jupyter}, $\mathtt{Matplotlib}$ \citep{Matplotlib07, Matplotlib16}, $\mathtt{NumPy}$ \citep{NumPy11, NumPy20}, $\mathtt{pandas}$ \citep{mckinney-proc-scipy-2010, reback2020pandas}, $\mathtt{PyMC}$ \citep{pymc}, $\mathtt{SciPy}$ \citep{2020SciPy-NMeth}, $\mathtt{showyourwork}$ \citep{Luger2021}}

\bibliography{bib}

\end{document}